\documentclass[reprint,superscriptaddress,showpacs,preprintnumbers,nofootinbib,amsmath,amssymb,floatfix,aps,prd,twocolumn,longbibliography]{revtex4-1}
\usepackage{graphicx}
\usepackage{subfigure}%
\usepackage{dcolumn}

\newcommand{\ve}[1]{\ensuremath{\mathbf{#1}}}
\newcommand{\n}[1]{\ensuremath{|\mathbf{#1}|}}
\newcommand{\nuC}{\ensuremath{^{12}_{\phantom{1}6}\textrm{C}(\nu_\mu,\mu^-)X}}
\newcommand{\ph}{\ensuremath{2p2h}}
\newcommand{\GENIE}{\textsc{genie}}
\newcommand{\vt}{\textsc{genie}$+\nu$T}
\newcommand{\GLoBES}{\textsc{gl}{\small o}\textsc{bes}}

\newcommand{\etal}{{\it et al.}}

\graphicspath{{figs/}}

\begin{document}


\title{Effect of the $2p2h$ cross-section uncertainties on an analysis of neutrino oscillations}

\author{Artur M. Ankowski}
\email{ankowski@vt.edu}
\affiliation{Center for Neutrino Physics, Virginia Tech, Blacksburg, Virginia 24061, USA}
\author{Omar Benhar}
\affiliation{INFN and Department of Physics,``Sapienza'' Universit\`a di Roma, I-00185 Roma, Italy}
\affiliation{Center for Neutrino Physics, Virginia Tech, Blacksburg, Virginia 24061, USA}
\author{Camillo Mariani}
\affiliation{Center for Neutrino Physics, Virginia Tech, Blacksburg, Virginia 24061, USA}
\author{Erica Vagnoni}
\affiliation{INFN and Dipartimento di Matematica e Fisica, Universit\`a di Roma Tre, Via della Vasca Navale 84, 00146 Rome, Italy}

\begin{abstract}
We report the results of a study aimed at quantifying the impact  on the oscillation analysis of the uncertainties associated with the description of the
neutrino-nucleus cross section in the two-particle--two-hole sector.
The results of our calculations, based on the kinematic method of energy reconstruction and carried out comparing two data-driven approaches, show that the existing discrepancies in the neutrino cross sections have a sizable effect on the extracted oscillation parameters,
particularly in the antineutrino channel.
\end{abstract}


\maketitle

The T2K Collaboration has recently reported two measurements of the inclusive cross section for charged-current (CC) muon-neutrino scattering off the hydrocarbon target, CH~\cite{Abe:2013jth,Abe:2014nox}. Being flux-averaged at different mean-energy values, the T2K results show the cross section as a function of neutrino energy with minimal dependence on nuclear models.

While the T2K data are lower by $\sim$20\% than the flux-averaged hydrocarbon result previously obtained by the SciBooNE Collaboration~\cite{Nakajima:2010fp}, with the difference exceeding the experimental uncertainties, they appear to be in good agreement with the expectations based on the \nuC{} cross section measured at higher energies by the NOMAD experiment~\cite{Wu:2007ab}.

At the kinematics of the T2K and SciBooNE experiments, momentum transfers $\ve q$ are typically large enough for neutrinos---probing the nuclear interior with the spatial resolution $\sim1/\n q$---to scatter off individual (bound) nucleons. On the other hand, the dominant contribution to the cross section comes from low energy transfers $\omega$, insufficient to produce pions, and the quasielastic (QE) mechanisms of interaction,
\begin{equation}\begin{split}
\nu_\ell+n&\rightarrow \ell^-+p,\\
\bar\nu_\ell+p&\rightarrow \ell^++n,
\end{split}\end{equation}
play the most important role.

In the past, CC QE processes were considered well understood theoretically and used to determine the flux normalization~\cite{Ahrens:1986ke}. Recently, however, it has become apparent that this is not the case to the extent required by precise oscillation experiments~\cite{Boyd:2009zz}. For example, while the CC QE cross sections of carbon reported by the MiniBooNE Collaboration~\cite{AguilarArevalo:2010zc,AguilarArevalo:2013hm} turn out to be higher than those of free nucleons, the corresponding NOMAD data~\cite{Lyubushkin:2008pe} show the cross-sections' reduction arising from  nuclear effects.
Although those puzzling discrepancies have received a great deal of theoretical interest, their interpretation is not fully established so far.

In particular, while a non-negligible role of CC QE reaction mechanisms involving more than one nucleon is now generally acknowledged, and important theoretical progress has been achieved~\cite{Benhar:2015ula}, an {\it ab initio} estimate of the corresponding cross sections is not yet available. As those multinucleon mechanisms involve predominantly two nucleons, hereafter we refer to them as two-particle--two-hole (\ph{}) processes.

For nuclear targets ranging from carbon to iron, a growing body of experimental evidence~\cite{Gran:2006jn,AguilarArevalo:2007ab,Mariani:2008zz,AguilarArevalo:2010zc,Abe:2014iza,Adamson:2014pgc} shows that \ph{} effects on the differential QE cross sections can be effectively accounted for by
increasing the value of the axial mass $M_A$, typically to $\sim$1.2 GeV, with respect to $M_A=1.03$ GeV extracted predominantly from deuterium measurements~\cite{Bernard:2001rs}. Note that as the axial mass is the cutoff parameter driving the axial form factor's dependence on $Q^2=\ve q^2-\omega^2$, its changes affect both the differential and total cross sections.

In this article, we discuss uncertainties of the \ph{} cross sections for carbon and quantify their effect on the oscillation analysis for an experimental setup similar to that of T2K~\cite{Abe:2011ks}. We consider a disappearance experiment running in both neutrino and antineutrino mode with the same flux~\cite{Huber:2009cw}, peaked at $\sim$600 MeV. To describe the ground-state properties of the target nucleus, we use the realistic spectral function (SF) of Ref.~\cite{Benhar:1994hw}. This approach allows an accurate estimate of QE scattering induced by one-nucleon currents, as shown by an extensive comparison to electron-scattering data in Ref.~\cite{Ankowski:2014yfa}. To account for an increase of the CC QE cross sections due to \ph{} processes, we use two data-driven phenomenological methods: (i) an increased value of the axial mass, yielding results consistent with the T2K~\cite{Abe:2013jth,Abe:2014nox}, NOMAD~\cite{Wu:2007ab,Lyubushkin:2008pe}, and MINERvA~\cite{Fiorentini:2013ezn,Fields:2013zhk} data, and (ii) the \ph{} estimate in the \GENIE{} Monte Carlo generator~\cite{Katori:2013eoa}, determined from the MiniBooNE data~\cite{AguilarArevalo:2010zc} and in agreement with the experimental cross sections extracted from SciBooNE~\cite{Nakajima:2010fp}.

\begin{figure*}
\centering
    {\includegraphics[width=0.44\textwidth]{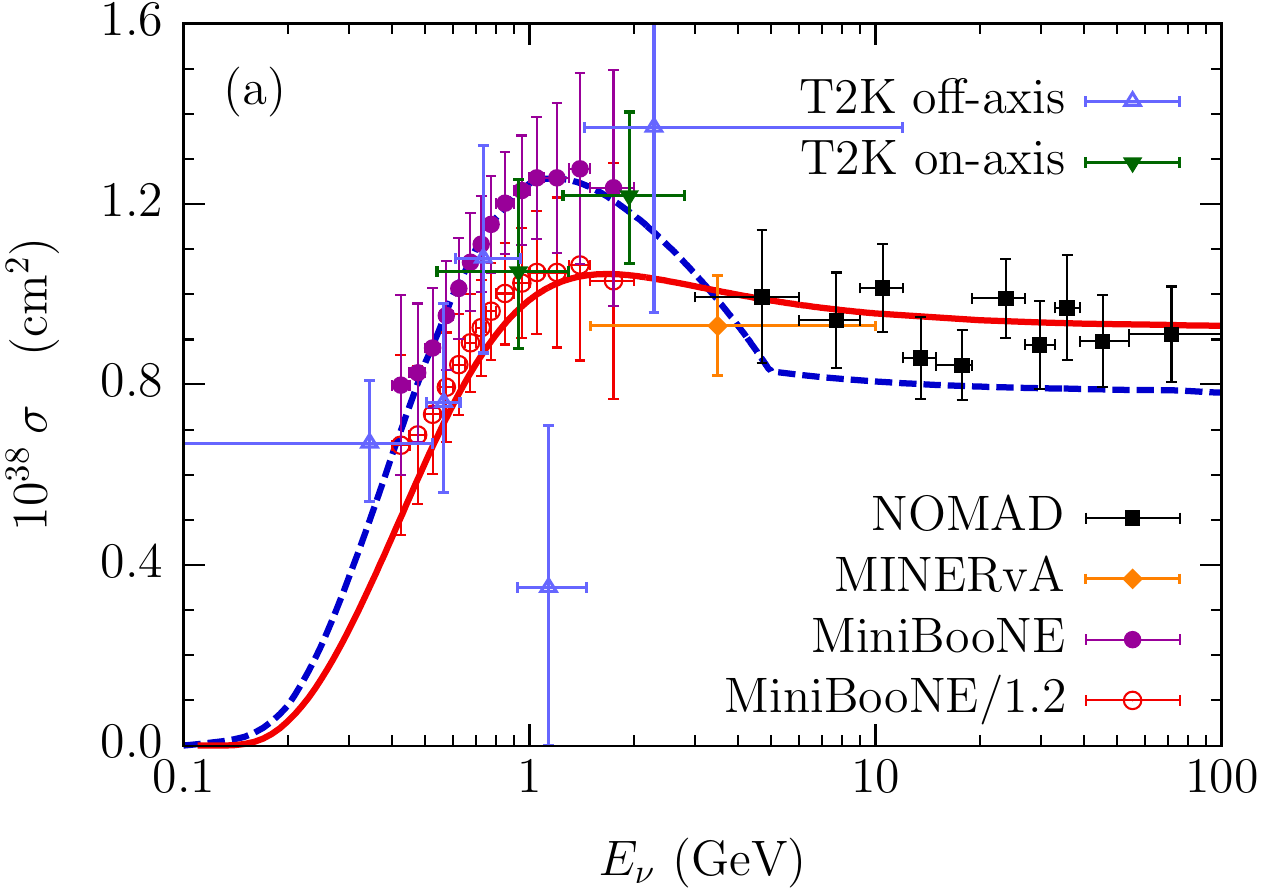}}
    \hspace{0.8cm}
    {\includegraphics[width=0.44\textwidth]{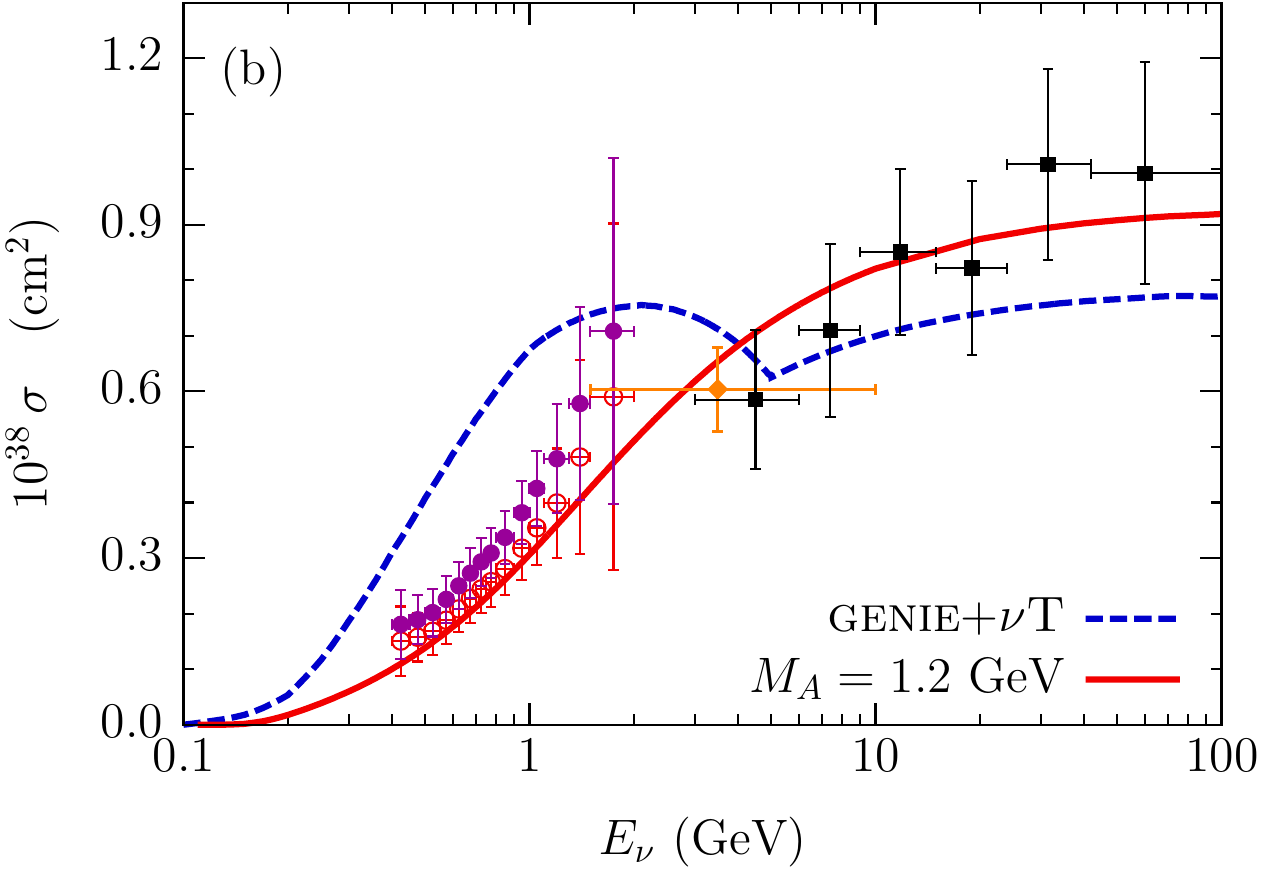}}
\caption{\label{fig:xsec_QE} CC QE (a) $\nu_\mu$ and (b) $\bar\nu_\mu$ cross sections. The results for carbon obtained using \vt{} (dashed lines) and the SF approach with $M_A=1.2$ GeV (solid line) are compared with the carbon data reported by the MiniBooNE~\cite{AguilarArevalo:2010zc,AguilarArevalo:2013hm} and NOMAD~\cite{Lyubushkin:2008pe} Collaborations and the hydrocarbon data extracted from the MINERvA~\cite{Fiorentini:2013ezn,Fields:2013zhk} and T2K~\cite{Abe:2015oar,Abe:2014iza} experiments. For comparison, the MiniBooNE data divided by 1.2 are also shown.}
\end{figure*}

We emphasize that although our study is performed for a setup similar---not identical---to that of T2K, it does not follow the analysis of that experiment.
For example, applying a generalization of the kinematic method of energy reconstruction~\cite{Ankowski:2015jya}, we include in the oscillation analysis events of \emph{all types}, instead of the CC QE event sample alone. The rationale for considering the T2K-like kinematics is its importance for the next generation of oscillation experiments~\cite{Abe:2015zbg,Acciarri:2016crz}.

Consequences of \ph{} effects for the CC QE cross sections have been analyzed within a few effective approaches. The calculations of Martini {\etal}~\cite{Martini:2009uj,Martini:2010ex,Martini:2011wp,Martini:2013sha}, based on the local Fermi gas model and the random-phase approximation (RPA), extend the treatment of multinucleon contributions to the electromagnetic responses of iron developed by Alberico {\etal}~\cite{Alberico:1983zg} to the case of neutrino interactions with carbon and to a broader kinematic region.

While employing the local Fermi gas model and the RPA scheme, the approach of Nieves \etal~\cite{Nieves:2011pp,Nieves:2011yp,Nieves:2013fr,Gran:2013kda} differs from that of Martini \etal{} by using effective interactions, the parameters of which were fixed in earlier studies of photon, electron, and pion scattering off nuclei.
At the MiniBooNE kinematics, the CC QE $\nu_\mu$ ($\bar\nu_\mu$) cross sections obtained by Nieves \etal{} are lower by $\sim$10\% ($\sim$15\%) with respect to those calculated by Martini \etal{}

To extend their superscaling approach and include the contributions of processes involving two-nucleon currents, Amaro \etal{}~\cite{Amaro:2010sd,Amaro:2011aa} and Megias \etal{}~\cite{Amaro:2013yna,Megias:2014qva} have previously estimated the \ph{} cross sections within the relativistic Fermi gas model accounting for the vector meson-exchange currents only. Recently, the efforts to also include the axial part in the response functions have been completed~\cite{Simo:2016ikv}.

In the Giessen Boltzmann-Uehling-Uhlenbeck transport model, the \ph{} contribution to the CC QE cross sections is obtained from a fit to the MiniBooNE data performed by Lalakulich \etal{}~\cite{Lalakulich:2012ac}, using a physically well-motivated  {\it ansatz}.

The \GENIE{} Monte Carlo generator~\cite{Andreopoulos:2009rq} simulates \ph{} events following the empirical procedure developed by Dytman~\cite{Katori:2013eoa}, based on the one derived for electron scattering in Ref.~\cite{Lightbody:1988}. The kinematics of the produced lepton is distributed according to the magnetic contribution to the elementary cross section and, as a consequence, turns out to be the same for neutrinos and antineutrinos. The \ph{} strength is set to decrease linearly for neutrino energy larger than
1 GeV and to vanish at 5 GeV, consistently with both the MiniBooNE~\cite{AguilarArevalo:2010zc} and NOMAD~\cite{Lyubushkin:2008pe} data. \GENIE{} is employed in data analysis by a number of neutrino experiments~\cite{Dytman:2011zza}, as well as in phenomenological estimates of the impact of nuclear effects on the determination of oscillation parameters, following the pioneering studies carried out by the authors of Ref.~\cite{Coloma:2013rqa}.

In this article, we analyze how the oscillation analysis may be affected by uncertainties in the description of  \ph{} contributions to the CC QE cross sections, comparing two estimates obtained from different approaches.
In the first case, we apply an effective value of the axial mass $M_A=1.2$ GeV to account for the modifications of the QE cross sections due to \ph{} reaction mechanisms in a purely phenomenological manner (``effective'' calculations).
In the second case, we add the \ph{} results obtained using \GENIE{} 2.8.0~\cite{Katori:2013eoa} to the QE calculations performed using the SF approach with $M_A=1.03$ GeV, as implemented in the $\nu$T package of additional modules~\cite{Jen:2014aja} (``\vt{}'' calculations).

\begin{figure*}
\centering
    {\includegraphics[width=0.44\textwidth]{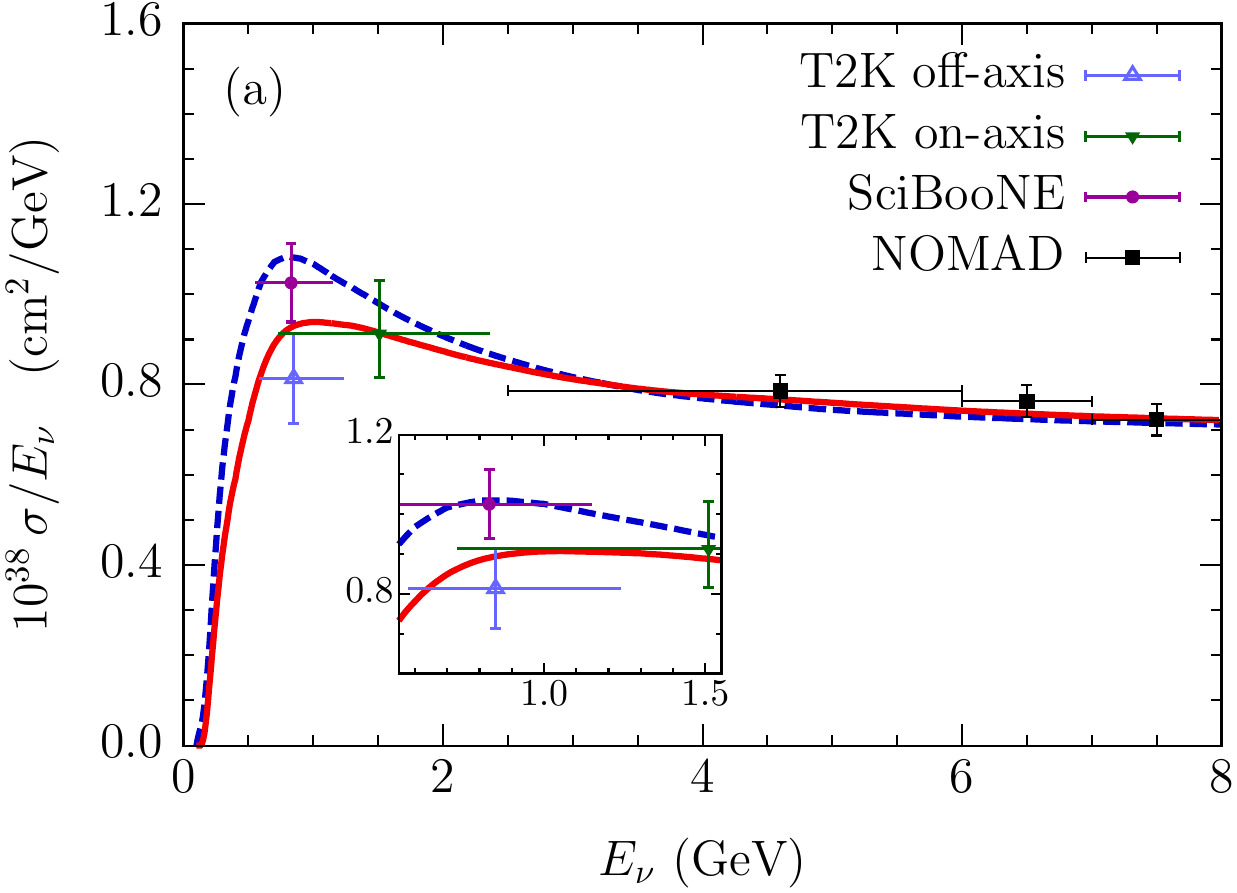}}
    \hspace{0.8cm}
    {\includegraphics[width=0.44\textwidth]{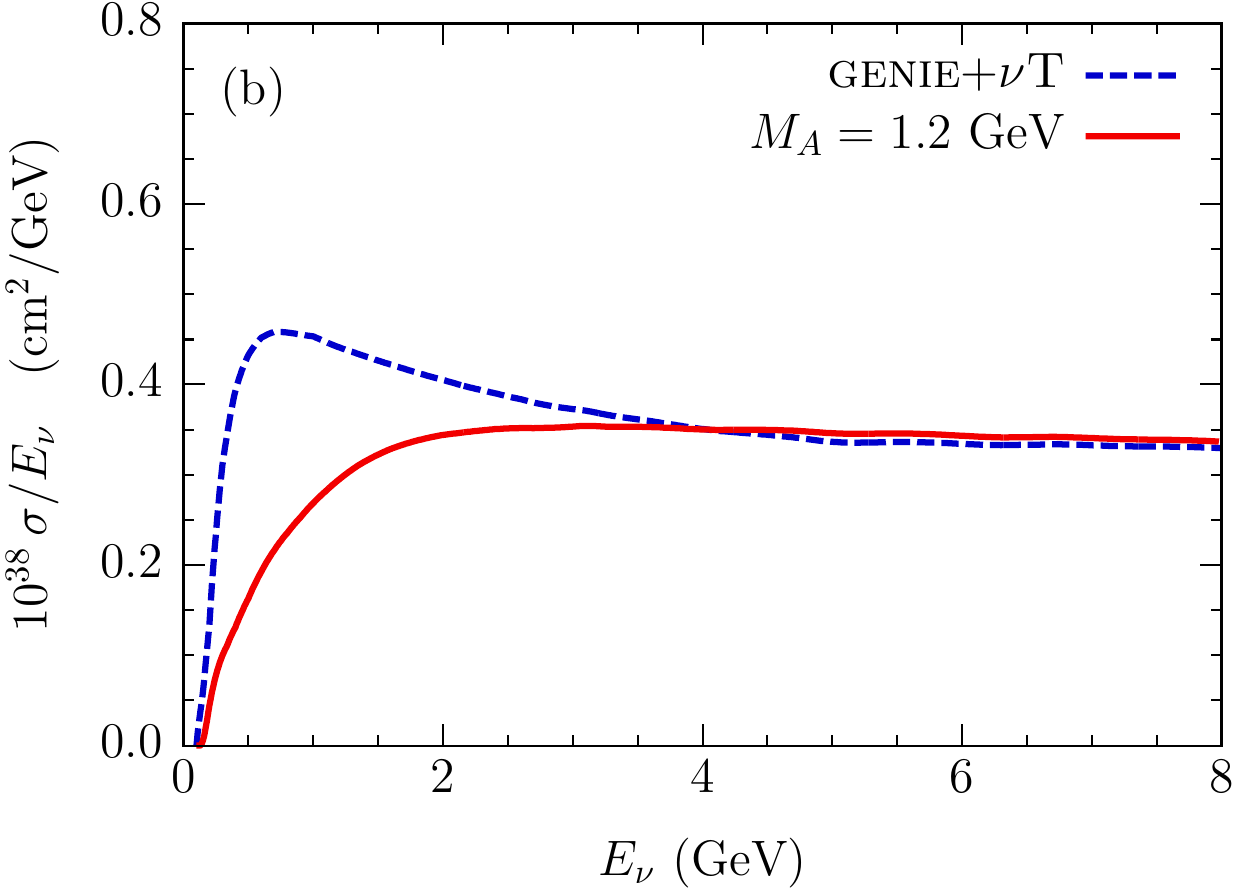}}
\caption{\label{fig:xsec_inc}Per-nucleon CC inclusive (a) $\nu_\mu$ and (b) $\bar\nu_\mu$ cross sections divided by neutrino energy, obtained using the QE contributions of Fig.~\ref{fig:xsec_QE}. The calculations for the carbon target (and for the hydrocarbon target in the inset) are compared with the carbon data extracted from the NOMAD~\cite{Wu:2007ab} experiment and the hydrocarbon flux-averaged measurements reported by the SciBooNE~\cite{Nakajima:2010fp} and T2K~\cite{Abe:2013jth,Abe:2014nox} Collaborations (the central energy values correspond to the mean energy in the detector). Note that antineutrino data are currently unavailable.}
\end{figure*}

The obtained total CC QE cross sections are compared to the experimental data in Fig.~\ref{fig:xsec_QE}. It clearly appears that the effective calculations are in good agreement with the NOMAD~\cite{Lyubushkin:2008pe} and MINERvA~\cite{Fields:2013zhk,Fiorentini:2013ezn} results  for both neutrinos and antineutrinos. They also reproduce the energy dependence of the MiniBooNE data~\cite{AguilarArevalo:2010zc,AguilarArevalo:2013hm}, but not their absolute normalization. To better illustrate this feature, we have divided the MiniBooNE cross sections by a factor of 1.2, consistent with the ratio of the detected to predicted events of $1.21\pm0.24$ reported from the first MiniBooNE analysis~\cite{AguilarArevalo:2007ab}.

While for neutrinos, the \ph{} contribution from \GENIE{} is in very good agreement with the MiniBooNE data, for antineutrinos it overestimates the experimental points, in spite of  being added to the SF
results obtained using $M_A=1.03$ GeV, which are too low to reproduce the cross sections from NOMAD~\cite{Ankowski:2012ei}.  Owing to their large uncertainties, the T2K CC QE data~\cite{Abe:2015oar,Abe:2014iza} cannot discriminate between the two calculations.

Adding the considered CC QE estimates to the cross sections for resonant, nonresonant, and coherent pion production from \GENIE{}, we have calculated the inclusive CC cross sections for carbon shown in
Fig.~\ref{fig:xsec_inc}. The two considered approaches turn out to be in good agreement with the NOMAD data~\cite{Wu:2007ab}, collected in the region dominated by pion production.

To compare to the T2K~\cite{Abe:2013jth,Abe:2014nox} and SciBooNE~\cite{Nakajima:2010fp} data, extracted for the hydrocarbon target, we have accounted for the contribution of free protons using the cross sections from \GENIE{}.
While the on-axis T2K data point~\cite{Abe:2014nox} does not distinguish the two approaches, the SciBooNE point~\cite{Nakajima:2010fp} clearly favors the \vt{} calculations and the off-axis T2K point~\cite{Abe:2013jth} shows a distinct preference for the effective calculations.

\begin{table}
\caption{\label{tab:osc_val}The oscillation parameters assumed in the analysis.}
\begin{ruledtabular}
\begin{tabular}{c c c c c c}
$\Delta m_{21}^2$ (eV$^2$) & $\Delta m_{31}^2$ (eV$^2$) & $\theta_{12}$ $(^\circ)$ & $\theta_{23}$ $(^\circ)$& $\theta_{13}$ $(^\circ)$& $\delta$ \\[3pt]
$7.50 \times 10^{-5}$ & $2.46 \times 10^{-3}$ & 33.48 & 42.30 &  8.50 & 0.0
\end{tabular}
\end{ruledtabular}
\end{table}


\begin{figure*}
\begin{center}
\includegraphics[scale=0.22]{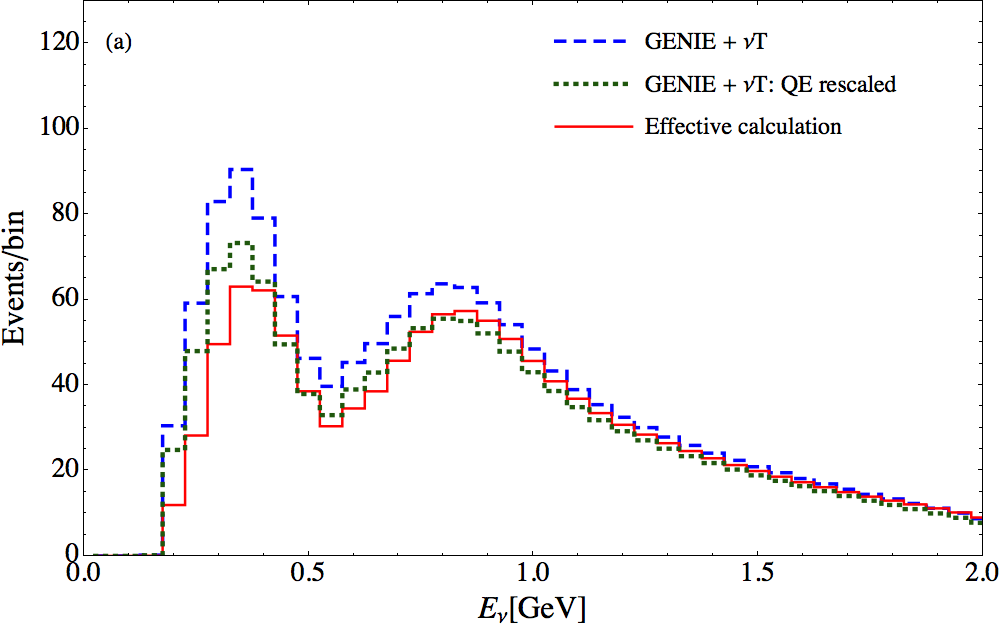}\hspace{0.8cm}
\includegraphics[scale=0.22]{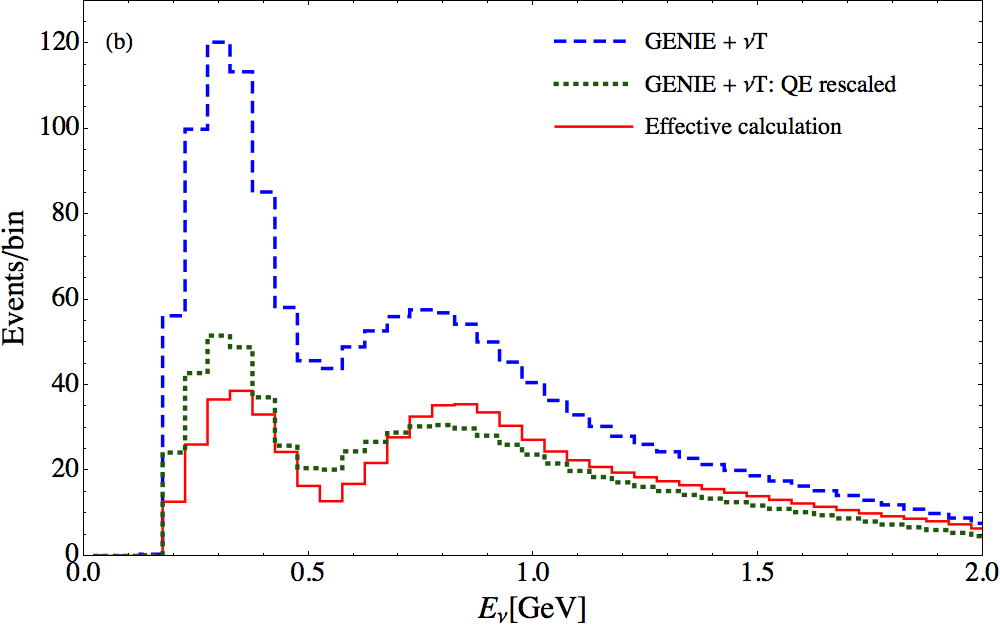}
    \subfigure{\label{fig:rate_numu}}
    \subfigure{\label{fig:rate_anumu}}
\caption{\label{fig:rate} (color online). Distribution of CC (a) $\nu_\mu$ and (b) $\bar\nu_\mu$ events
  in the far detector as a function of the reconstructed energy, obtained within the \vt{} and effective calculations. For comparison, we also show the \vt{} results with the unoscillated QE event rates rescaled to those of the effective calculations.}
\end{center}
\end{figure*}

\begin{figure*}
\begin{center}
\includegraphics[width=0.42\textwidth]{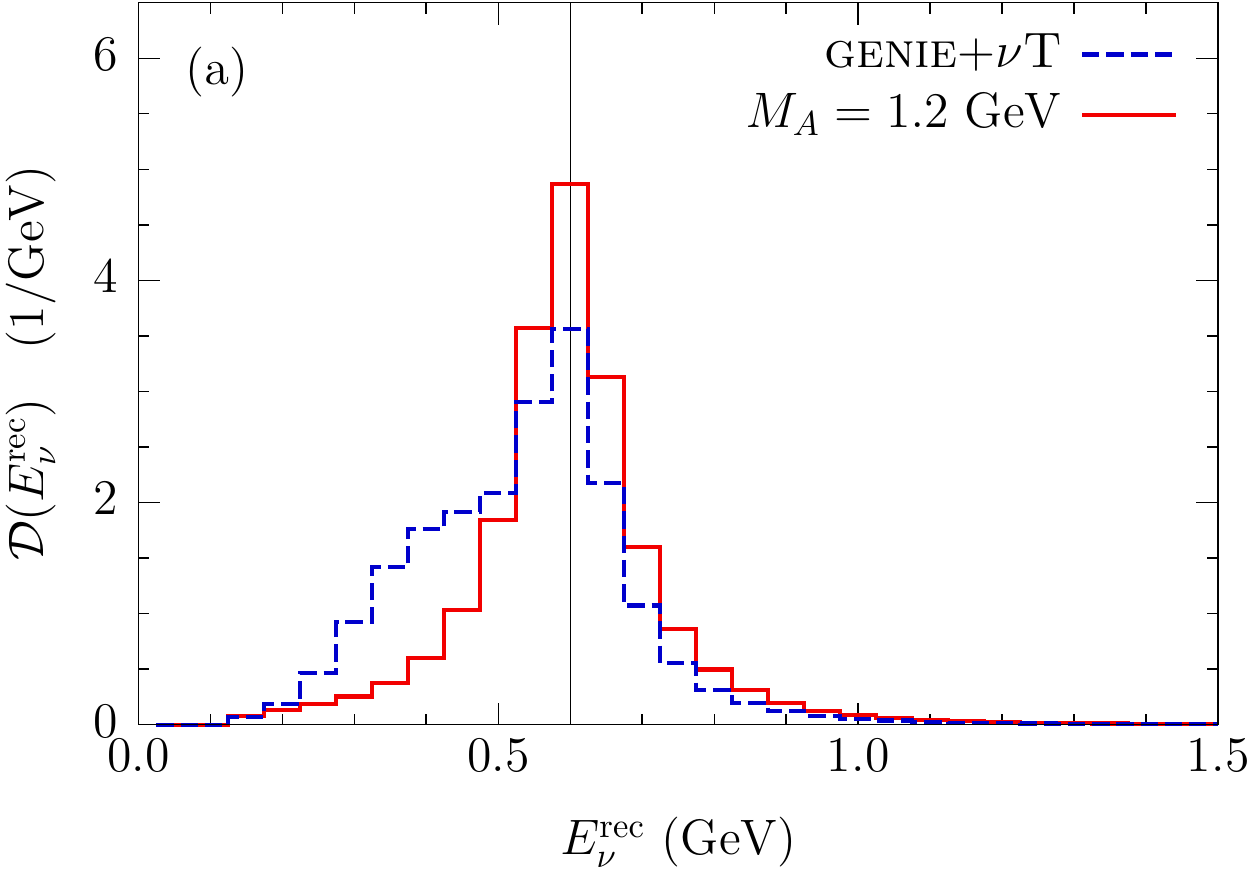}
\hspace{0.8cm}
\includegraphics[width=0.42\textwidth]{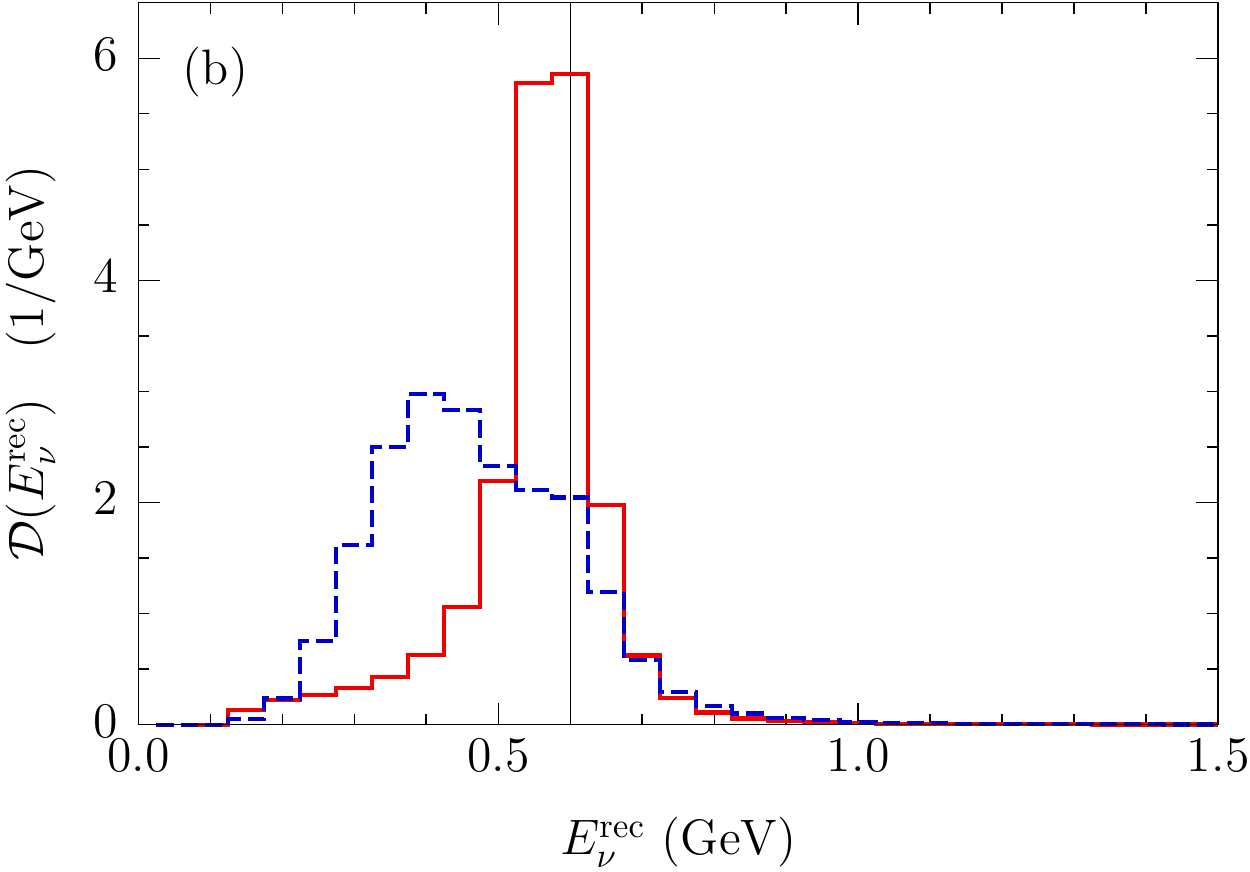}
    \subfigure{\label{fig:recE_numu}}
    \subfigure{\label{fig:recE_anumu}}
\caption{\label{fig:recE} (color online). Reconstructed energy distributions of CC QE (a) $\nu_\mu$ and (b) $\bar\nu_\mu$ events with any number of nucleons calculated at $E_\nu = 0.6$ GeV. The dashed (solid) lines represent the results obtained using the \vt{} (effective) approach.}
\end{center}
\end{figure*}

\begin{figure*}
\begin{center}
\includegraphics[scale=0.2]{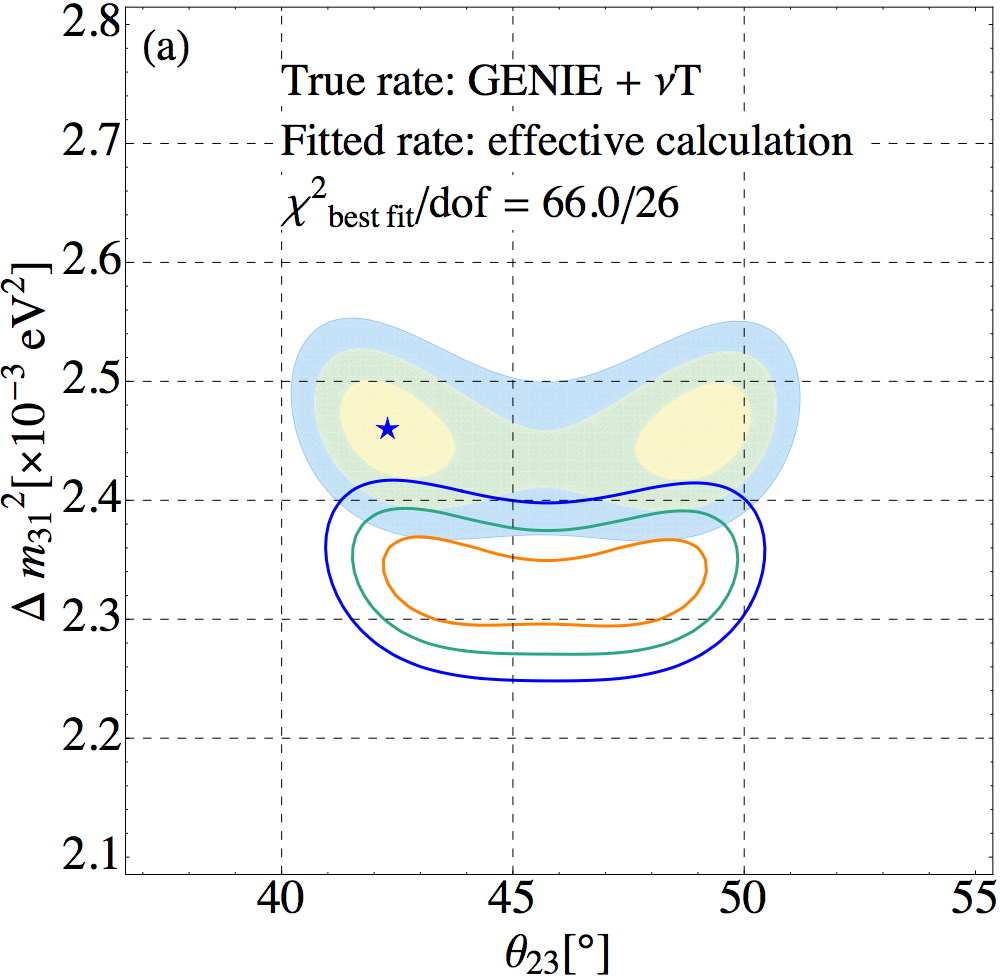}
\hspace{0.8cm}
\includegraphics[scale=0.2]{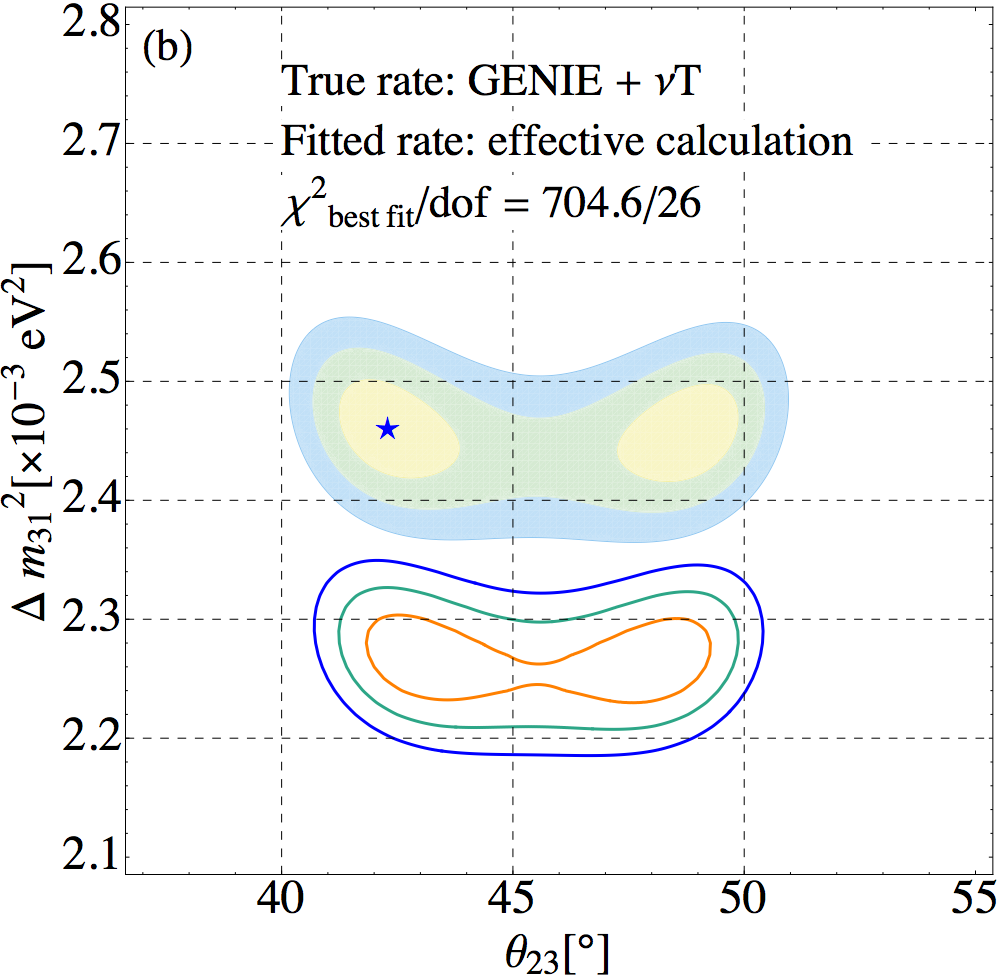}
    \subfigure{\label{fig:conf_numu}}
    \subfigure{\label{fig:conf_anumu}}
\caption{\label{fig:conf} (color online). Confidence regions in the $(\theta_{23}, \Delta m^2_{31})$ plane for the true (a) $\nu_\mu$ and (b) $\bar\nu_\mu$ event rates from \vt{} at the 1, 2 and 3$\sigma$ C.L. The shaded areas (solid lines) correspond to the fitted rates from the \vt{} (effective) calculations.}
\end{center}
\end{figure*}

The puzzling difference between the T2K and SciBooNE data---interesting in its own right---has important consequences for neutrino-oscillation studies. We discuss them for the $\nu_\mu$ and $\bar\nu_\mu$ disappearance analysis of an experiment using an off-axis ($2.5^\circ$) beam peaked at $\sim$600 MeV~\cite{Huber:2009cw}. The near (far) detector with a fiducial mass of 1.0 (22.5) kton is located at a distance of 1 (295) km from the neutrino source.

We adopt the kinematic method of energy reconstruction, applying it to all event types as in Ref.~\cite{Ankowski:2015jya}. As neutral-current background is expected not to play an important role, we do not take it into account. Our analysis employs \GLoBES{}~\cite{Huber:2004ka,Huber:2007ji,Coloma:2012ji} and is based on $\sim$6000 unoscillated events with reconstructed energies between 0.3 and 1.7 GeV, in both the neutrino and antineutrino modes. The oscillation-parameter values assumed as the true ones are detailed in Table~\ref{tab:osc_val}. Implementing $\chi^2$, we apply a $20\%$ systematic uncertainty of the shape (normalization), bin-to-bin uncorrelated (correlated).

In our analysis, the true event rates are simulated using the \vt{} calculations, and the fitted rates are obtained for both considered approaches over a range of atmospheric oscillation parameters, $\theta_{23}$ and $\Delta m_{31}^2$.
Having determined the minimal $\chi^2$ value, $\chi^2_\textrm{best-fit}$, the confidence regions are found from the condition
\begin{equation}
\label{eq:chi2}
\Delta \chi^2 (\theta_{23}, \Delta m^2_{31}) \equiv \chi^2 (\theta_{23}, \Delta m^2_{31}) - \chi^2_\textrm{best-fit} < l,
\end{equation}
where $l = 2.30$, 6.18, and 11.83 for the 1, 2, and $3\sigma$ confidence level, respectively.

Before discussing the oscillation results, it is illustrative to compare the reconstructed energy distributions for muon neutrinos and antineutrinos obtained from the \vt{} and effective calculations. As shown in Fig.~\ref{fig:rate}, the differences between the two cross-section estimates translate into differences between the oscillated event rates in the far detector, with the discrepancies being particularly severe in the case of antineutrinos.

In addition to the total event numbers, the two approaches yield clearly different distributions of reconstructed energy,
as shown in Fig.~\ref{fig:recE} for the true energy $E_\nu=0.6$ GeV and in the Supplemental Material for $0.2\leq E_\nu\leq2.0$ GeV~\cite{SupplementalMaterial}. 
While in the effective calculations, \ph{} processes enhance the low-energy tails of the distributions, in the \vt{} approach, they also produce additional bumps, corresponding to the reconstructed energy $\sim$0.4 GeV at the kinematics of Fig.~\ref{fig:recE}. In the antineutrino case, for $E_\nu\lesssim 1.4$ GeV the strength of these \ph{} bumps turns out to be larger than that of the QE ones, located at $E^\textrm{rec}_\nu\simeq E_\nu$. The observed differences in the reconstructed energy distributions have important consequences for the oscillation analysis.

The obtained confidence regions are shown in Fig.~\ref{fig:conf}. The shaded areas represent the results for the \vt{} fitted rates, and the solid lines correspond to the fitted rates from the effective calculations. The high values of $\chi^2_\textrm{best-fit}$ per degree of freedom, given in Fig.~\ref{fig:conf}, clearly indicate that the differences between the two considered approaches are too large to be neglected in a precise oscillation analysis. We have verified that this observation holds true even when the normalization of the QE event sample, with any number of nucleons, is treated as arbitrary. Therefore, the observed effect can be traced back to the shape discrepancies displayed in Figs.~\ref{fig:xsec_QE} and~\ref{fig:recE}, which appear to be especially large for antineutrinos.
In particular, as for antineutrinos in the relevant $E_\nu$ region the reconstructed energy distributions in the effective and \vt{} approaches are peaked at different values, the extracted $\Delta m_{31}^2$ is subject to larger bias for antineutrinos than for neutrinos.

In summary, we have studied the impact of discrepancies between experimental cross sections on neutrino-oscillation analysis, adopting the kinematic method of energy reconstruction. We have compared two data-driven approaches focusing on the 1-GeV energy region and shown that the differences between them have a sizable effect on the resulting oscillation parameters, especially in the antineutrino channel.

In view of these findings, improving the precision of the neutrino and antineutrino cross sections will be of great importance for future oscillation studies. Such progress will require new experimental data
for energies $\sim$1 GeV, as well as an  improvement in the understanding of systematic uncertainties, which would allow the tensions between existing measurements to be significantly alleviated.

Because the description of final-state hadrons involves larger uncertainties than those associated with leptons, the conclusions of this article are expected to also apply to the calorimetric method of energy reconstruction and are, therefore, relevant to the next generation of long-baseline oscillation measurements, such as the Deep Underground Neutrino Experiment~\cite{Acciarri:2016crz}, aimed at  determining the charge-parity violating phase and at verification of the three-neutrino paradigm.

\section*{Acknowledgments}
We would like to express our gratitude to Chun-Min Jen for generating the events used in this analysis and for extracting  the cross sections from \GENIE{}. E.V. is indebted to Davide Meloni for many
useful discussions. Preparing the manuscript, we also benefited from the comments of Pilar Coloma and Steve Dytman. The work of A. M. A. and C. M. was supported by the National Science Foundation under Grant No. PHY-1352106.


%

\end{document}